\documentclass[pdflatex,sn-nature]{sn-jnl}

\usepackage{graphicx}%
\usepackage{multirow}%
\usepackage{amsmath,amssymb,amsfonts}%
\usepackage{amsthm}%
\usepackage{mathrsfs}%
\usepackage[title]{appendix}%
\usepackage{xcolor}%
\usepackage{textcomp}%
\usepackage{manyfoot}%
\usepackage{booktabs}%
\usepackage{algorithm}%
\usepackage{algorithmicx}%
\usepackage{algpseudocode}%
\usepackage{listings}%

\theoremstyle{thmstyleone}%
\theoremstyle{thmstyletwo}%

\theoremstyle{thmstylethree}%

\begin{document}

\title[Understanding and predicting trends in adsorption energetics on monolayer transition metal dichalcogenides]{Understanding and predicting trends in adsorption energetics on monolayer transition metal dichalcogenides}

\author[1]{\fnm{Brian H.} \sur{Lee}}

\author[2]{\fnm{Jameela} \sur{Fatheema}}

\author[1,2]{\fnm{Deji} \sur{Akinwande}}

\author*[1,3]{\fnm{Wennie} \sur{Wang}}\email{wwwennie@che.utexas.edu}

\affil[1]{\orgdiv{Texas Materials Institute}, \orgname{The University of Texas at Austin}}

\affil[2]{\orgdiv{Chandra Family Department of Electrical and Computer Engineering}, \orgname{The University of Texas at Austin}}

\affil[3]{\orgdiv{McKetta Department of Chemical Engineering}, \orgname{The University of Texas at Austin}}

\abstract{Two-dimensional (2D) transition metal dichalcogenides (TMDs) have emerged as promising candidates for non-volatile resistive switching (NVRS) due to their atomic-scale thickness, enabling high-density integration and low energy consumption. This study explores how metal adatom adsorption and desorption modulate resistivity in these materials. By examining a range of transition-metal adsorbates on MoS$_2$, MoSe$_2$, WS$_2$, and WSe$_2$, we uncover material-specific trends in adsorption energies that correlate with electronic and atomic structure descriptors. These trends are captured by simple models that elucidate how adsorption influences the formation and dissolution of point defects, a key mechanism in NVRS. Our results highlight consistent periodic behavior across TMDs and demonstrate that adsorption energy is a useful predictor of switching energy. This understanding provides insight into atomic-scale switching phenomena and offers design principles for selecting TMD-adsorbate pairs to optimize NVRS device performance.}

\maketitle
\begin{center}
    \textit{This is the accepted manuscript of the article published in \textbf{npj 2D Materials and Applications} (2025), DOI: \href{https://doi.org/10.1038/s41699-025-00579-9}{10.1038/s41699-025-00579-9}.}
\end{center}

\section{Introduction}\label{sec1}

Diversification of memory and storage technologies and improvements of 10$\times$-100$\times$ in capacity and energy efficiency are required, necessitating novel materials platforms \cite{ishimaru_future_2019,christensen2022RoadmapNeuromorphic2022}. 
The drive towards smaller and more energy-efficient electronic devices has heightened interest in two-dimensional (2D) materials ~\cite{sangwanNeuromorphicNanoelectronicMaterials2020, liResistiveMemoryDevices2024, rehmanDecade2DmaterialsbasedRRAM2020, lemme_2d_2022}. 
This interest stems from the unique electronic, optical, and scalability properties of the low-dimensional materials \cite{mounet_two-dimensional_2018, splendiani_emerging_2010, wang_electronics_2012, chhowalla_chemistry_2013, chhowalla_two-dimensional_2016}. 
Two-dimensional materials exhibiting non-volatile resistive switching (NVRS) \cite{shi_electronic_2018, yan_vacancyinduced_2019, ge_library_2021} have demonstrated high yields at wafer-scale \cite{chenWaferscaleIntegrationTwodimensional2020}, low cycle-to-cycle and device-to-device variability \cite{huhMemristorsBased2D2020}, and fast switching times \cite{wuThinnestNonvolatileMemory2019, teja_nibhanupudi_ultra-fast_2024},  making them competitive with existing and novel memory and storage technologies and neuromorphic computing applications \cite{lanzaMemristiveTechnologiesData2022}.
As a result, resistive switching based on atomically-thin 2D materials, including the transition metal dichalcogenides (TMDs) \cite{chowdhury_progress_2020, kwon_memristive_2022, obrien_process_2023, wu_atomristors_2020}, are increasingly viewed as promising materials platforms to address the pressing challenges of achieving higher density integration and enhanced energy efficiency in next-generation computing technologies.
How the switching characteristics of memristors depend on the switching and electrode materials is not fully understood.
In lateral configurations, where electrodes are placed on the same plane as the TMD layer, Sangwan et al. \cite{sangwan_gate-tunable_2015} demonstrated that gate-tunable memristive behavior in monolayer MoS$_2$ devices allows field-control of switching through electrostatic gating and operates via migration of grain boundaries via bias sweeps in the tens of volts.
Vertical memristors with electrodes positioned above and below the TMD layer exhibit distinct out-of-plane switching dynamics. 
Xu \textit{et al}. \cite{xu_vertical_2019} reported vertically stacked memristors using graphene/MoS$_2$/graphene structures, which provided excellent flexibility and endurance and switching voltages in the 100 milliVolt range. 
Kumar \textit{et al}. \cite{kumar_vertically_2019} further explored vertically aligned MoS$_2$ layers with metal electrodes, demonstrating that this configuration leads to enhanced on/off ratios and retention times due to improved layer-to-layer coupling.
Previous studies have achieved artificial synapses with combining different TMD layers or other 2D materials to form junctions ~\cite{bessonov_layered_2015,huh_synaptic_2018}, but these studies focused on imitating switching characteristics of biological synapses and device integration rather than a materials understanding.

Several mechanisms for resistive switching have also been proposed, including the migration of extended defects\cite{sangwan_gate-tunable_2015}, the generation of vacancies  \cite{yan_vacancyinduced_2019}, the formation of a conductive filament or bridge~\cite{xu_vertical_2019, zhao_current_2020, li_intrinsic_2021,mitra_atomistic_2024}, the migration of defect ions originating from the electrode~\cite{papadopoulos_ion_2022}, and relatedly, the adsorption and desorption of a metal adatom defect on a pre-existing chalcogen vacancy~\cite{ge_atomristor_2018, ge_library_2021,anvari_nature_2024,boschetto_non-volatile_2023,li_resistive_2023}. 
Thus far, there is a huge variability in reported switching characteristics in low-dimensional materials~\cite{rehmanDecade2DmaterialsbasedRRAM2020,duanLowPowerMemristorBased2022,zhouEmergingTwodimensionalMaterials2022}.
For example, MoS$_2$ with gold electrodes has been reported to have a higher switching ON/OFF ratio compared to WSe$_2$ in experiment~\cite{ge_library_2021} but the opposite trend is predicted from theory~\cite{li_resistive_2023}, highlighting a critical need to understand the switching and electrode materials and their defect chemistry.
Indeed, the plethora of NVRS mechanisms simultaneously highlights the prevalence of resistive switching in low-dimensional materials and the gaps in knowledge of the atomic underpinnings in resistive switching mechanisms at the material level. 

A materials-based understanding will help pave the way for exploring other transition metals, provide broader insights into optimizing memristive performance, and help in identifying new electrode and switching material combinations.
Several important computational studies have been made to understand defects in 2D materials and in resistive switching, including the generation~\cite{huangUnveilingComplexStructureproperty2023,gjerdingRecentProgressComputational2021,bertoldoQuantumPointDefects2022,davidssonAbsorptionAdsorptionHighthroughput2023} and analyses~\cite{frey_machine_2020} of several databases.
However, these databases and studies do not yet cover the space of defects relevant to resistive switching, e.g., the complex between an impurity and existing native point defect. 
Furthermore, to the best of our knowledge, very few limited systematic relationships have been determined for understanding the energetics of defects in low-dimensional materials.
In this study, we demonstrate that insights from electrochemistry can be directly applied to understand the defect energetics of materials relevant to resistive switching applications.

\begin{figure}[H]
    \centering
    \includegraphics[width=0.75\textwidth]{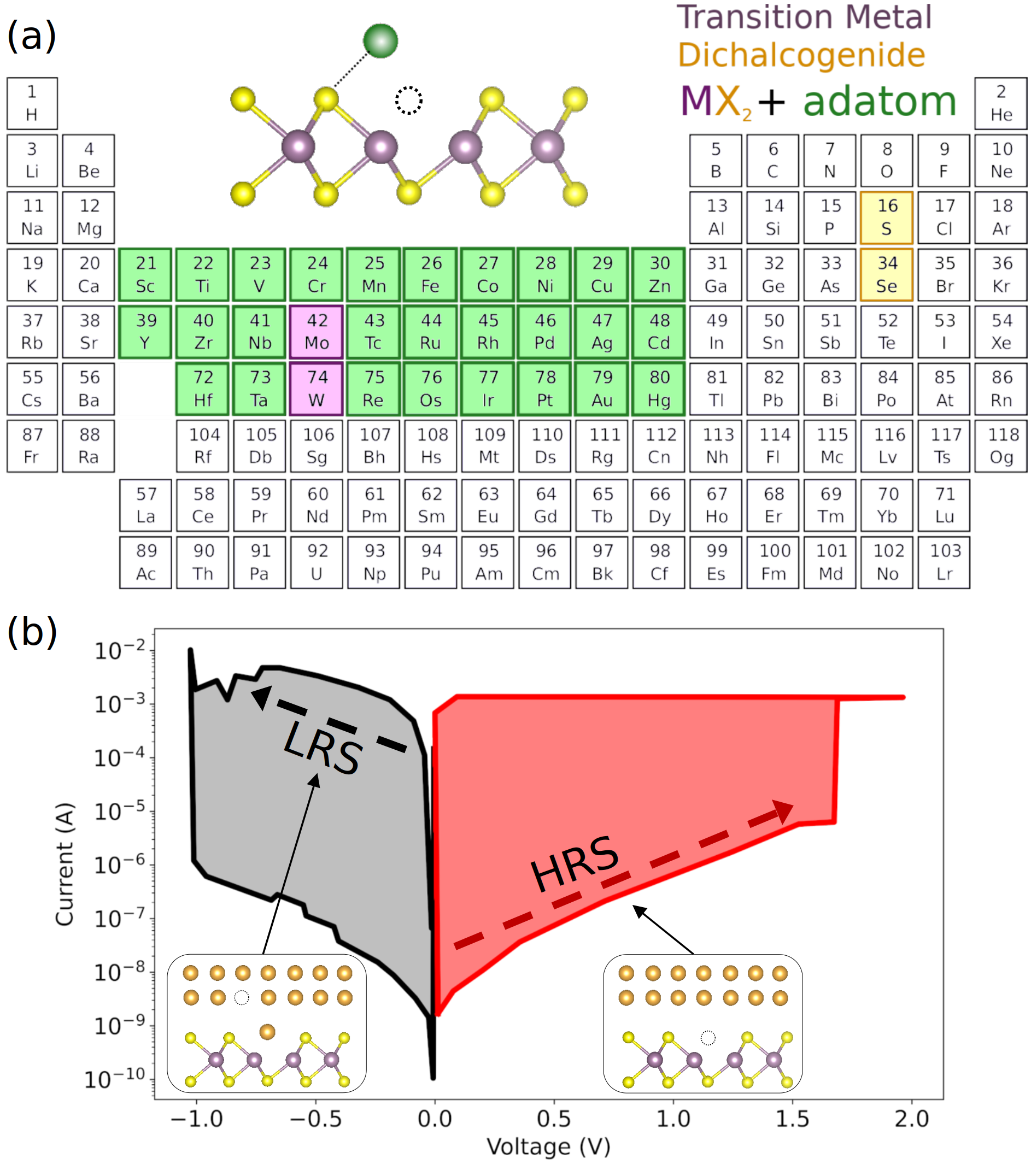}
    \caption{(a) Periodic table highlighting the elements studied in this work. Adatoms investigated for adsorption on transition metal dichalcogenides (TMDs) are marked in green. The transition metal (M) and chalcogen (X) elements in the MX$_2$ configuration are highlighted in purple and yellow, respectively. (b) Representative current-voltage (I-V) curve demonstrating resistive switching behavior in a MoS$_2$-based memristor device. The high resistance state (HRS) is indicated in red, corresponding to the system before adatom adsorption, while the low resistance state (LRS) is shown in black, corresponding to the state after adatom adsorption. Adapted from Ref~\cite{ge_library_2021} with permission. Copyright 2021 John Wiley and Sons.}
    \label{figure_overview}
\end{figure}

In this study, we utilize first-principles calculations based on density functional theory\cite{kohnSelfConsistentEquationsIncluding1965,hohenbergInhomogeneousElectronGas1964} to better understand the defect energetics in the resistive switching mechanism of monolayer transition metal dichalcogenides (TMDs).
We focus on resistive switching as mediated by the formation and dissolution of point defects, namely the adsorption and desorption of a (neutral) metal adatom that originates from the electrode material on the surface of the TMD (Figure~\ref{figure_overview}). 
Such defects are readily expected to form and contribute to resistive switching, as observed computationally~\cite{anvari_nature_2024} and experimentally~\cite{hus_observation_2021}.
In Section~\ref{sec_ads_trends}, we discuss trends observed in computed adsorption energies for representative monolayer TMDs (MoS$_2$, MoSe$_2$, WS$_2$, WSe$_2$) in order to extract meaningful materials descriptors.
In order to rationalize trends in adsorption energies, we introduce a model for predicting the adsorption energies of transition metal adsorbates on TMDs, including ones based on the \textit{d}-band center model inspired by the study of catalysts in Section~\ref{sec_dband_ads}.
We then present a framework for connecting the energetics of adsorption with the switching energy, defined as the energy required to change the resistive state of the device, inspired by the dissociation-diffusion-adsorption model \cite{ge_library_2021} in Section~\ref{sec_DDA}. 
Finally, we discuss ramifications of the study on considerations for materials selection in novel memory and computing technologies based on resistive switching mediated by point defects (Section~\ref{sec_discussion}).

Ultimately, this study seeks to address several key questions: 
Which adsorbates can induce NVRS? 
What dictates the adsorption energy and the switching energy? 
Are there discernible trends among potential materials candidates undergoing switching or for the electrode? 
Our study reveals that there are broad, underlying material trends in the defect energetics, which are correlated with the resistive switching behavior of the device. 
We utilize theoretical models such as the \textit{d}-band center theory, charge analysis, chemical bonding analysis, and compressed sensing techniques to rationalize trends in adsorption energetics. 
This study demonstrates that physical models of adsorption can be used to understand trends in energetics related to the switching energy in neuromorphic devices.
Notably, unlike conventional applications of the \textit{d}-band model involving organic adsorbates, we demonstrate that such adsorption physics can also apply to transition metal adsorbates.
Our approach simplifies the modeling of adsorption energy from existing models~\cite{frey_machine_2020}, which will provide a valuable tool for materials-driven design. 
By addressing these questions, this study aims to accelerate the identification of materials for memristive devices based on TMDs and other 2D materials.

\section{Results}\label{sec_results}
We now present results on the trends of adsorption energies and their relation to the switching energies. 
Results in the main text are mainly presented for MoS$_2$. 
Corresponding results for MoSe$_2$, WS$_2$, and WSe$_2$ are provided in the Supplementary Information (SI). 

\subsection{Trends in the adsorption energy}
\label{sec_ads_trends}
We examine the adsorption energies of different transition metal adatoms on TMD monolayers, revealing the underlying principles of their interactions. 
In particular, we focus on the adsorption of transition metal adatoms onto chalcogen vacancies, an intrinsic defect present in MoS$_2$, MoSe$_2$, WS$_2$, and WSe$_2$ samples \cite{kim_experimental_2022,anvari_nature_2024}. 
Our computed adsorption energies align well with prior computational studies.
For instance, the adsorption energy of gold onto a sulfur vacancy is found to be -2.64 eV, which is within $\pm$0.1 eV of values reported in the literature with slight variations arising from factors such as \textit{k}-point grid, supercell size, and choice of chemical potential \cite{miralrio_electronic_2018, frey_machine_2020, ge_library_2021}.

\begin{figure}[H] 
    \centering
    \includegraphics[width=0.95\linewidth]{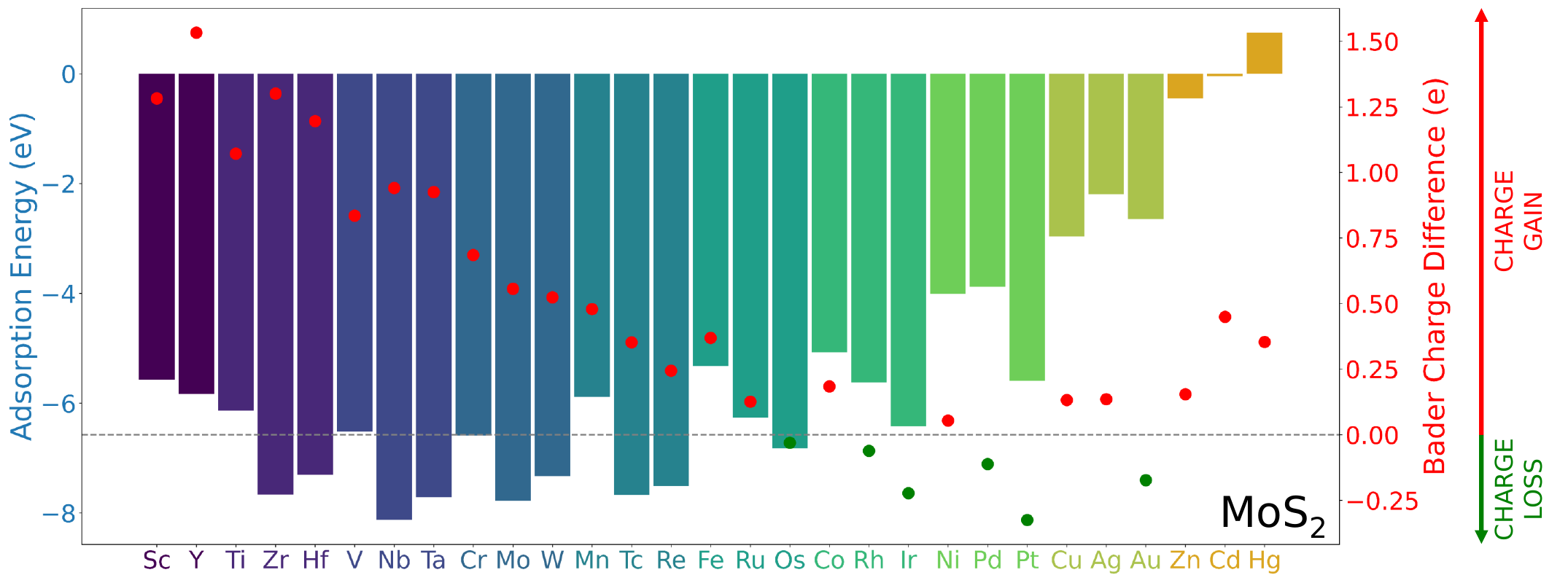}
    \caption{Adsorption energies (left, blue axis) of various transition metal adsorbates on a single sulfur vacancy in MoS$_2$. Adsorbates in each plot are sorted and colored by their periodic group. The right axis displays the Bader charge gain (green) or loss (red) of the adsorbate relative to its neutral atomic state, with negative values indicating charge gain and positive values indicating charge loss.}
    \label{figure_MoS2_ads}
\end{figure}

Figures~\ref{figure_MoS2_ads} and Supplementary Figure 6 show the computed adsorption energies for various 3$d$, 4$d$, and 5$d$ transition metal adsorbates and the corresponding change in Bader charge compared to the neutral atom. 
Bader charge analysis is a method used to partition the electronic charge density by assigning charges to atoms based on the topology of the electron density. 
This allows us to quantify how much charge is gained or lost by an atom during adsorption. 
When determining the adsorption energies of the different transition metal adsorbates onto the different monolayer TMDs, a trend can be observed. 
The relative order of adsorption energies for adsorbates remains consistent across different TMDs, with each adsorbate displaying a similar trend in adsorption strength on each substrate.
For example, on MoS$_2$, the adsorption energies are approximately -2.64 eV for gold, -2.19 eV for silver, and -2.96 eV for copper. 
This trend, where silver has the weakest adsorption energy, gold occupies an intermediate position, and copper has the strongest adsorption energy, is consistent across other TMDs. 
This indicates a minimal dependence on the identity of the cation or anion in the TMD (MoS$_2$, MoSe$_2$, WS$_2$, WSe$_2$ are all in the semiconducting 2H phase). 

As shown in Figure~\ref{figure_MoS2_ads}, adsorption energy and Bader charge gain or loss are loosely inversely correlated. 
Transition metal adsorbates that exhibit greater charge loss generally show stronger adsorption energies.
The earliest transition metals (Sc, Y) exhibit weaker adsorption energies than anticipated ($E^{\text{Sc}}_{ads}=-5.57$~eV, $E^{\text{Y}}_{ads}=-5.83$~eV) compared to later transition metal adatoms that lose less charge but have stronger adsorption interactions (e.g., $E^{\text{Ti}}_{ads}=-6.13$~eV, $E^{\text{Zr}}_{ads}=-7.66$~eV, $E^{\text{Hf}}_{ads}=-7.31$~eV).
However, the charge state of the adsorbate is only one aspect of the overall interaction.
Notable exceptions include osmium (Os), which exhibits minimal charge gain but still has a strong adsorption energy. 

When a chalcogen vacancy is present, the cation in the TMD becomes under-coordinated, leaving two excess electrons that alter the electronic environment of the transition metal adsorbate. 
This altered environment mediates the interaction strength between the adsorbate and the substrate, which depends not only on charge loss or gain but also on the adsorbate's ability to hybridize effectively with the substrate.

These trends in adsorption energy can be partially explained by the variation in valence electron configurations among adsorbates. 
Metals like Zn, Cd, and Hg, which have fully filled \textit{d}-orbitals and \textit{s}-orbitals, exhibit the weakest adsorption energies and, in some cases, even positive adsorption energies. 
A positive adsorption energy in this context indicates that these metals do not readily adsorb onto the vacancy. Their stable electron configurations result in lessened interaction with the unsaturated Mo atom, as these metals do not readily donate or accept electrons. 
Similarly, metals such as Ag, Au, and Cu, which have nearly filled \textit{d}-orbitals, also show relatively weak adsorption energies. Although Cu is not a noble metal, its nearly filled \textit{d}-orbital leads to weaker interactions with the TMD surface, similar to the behavior of Ag and Au.

In contrast, metals with partially filled \textit{d}-orbitals, such as Fe, typically exhibit intermediate adsorption energies.
The overall trend of stronger adsorption energies continues when examining the early transition metals, which have mostly unfilled \textit{d}-orbitals.
There is also an apparent periodicity in the adsorption energies with the extent of Bader charge loss or gain within each group. 
The periodicity in adsorption energy is altered when charge loss transitions into charge gain, which is notable for adsorbates such as Os, Ir, Pt, and Au.
As discussed in Section~\ref{sec_dband_ads}, multiple factors play a role in the adsorption energy, such as electron filling, orbital hybridization, and orbital geometry.

We note that the adsorption energy will vary depending on the value of the chemical potential, which will depend on experimental and device conditions. 
Thus, we look at two limits of the chemical potential where the adatom is referenced to an isolated adatom and to the monatomic metal bulk.
For the adsorption energies in Figure~\ref{figure_MoS2_ads}, the chemical potential is chosen as the isolated adatom, assuming dissociation of the adatom from the electrode has already occurred. 
When instead using the chemical potential referenced to the monatomic metal bulk, the adsorption energies can vary up to 10.93 eV, as seen in the case of tungsten (W), which had one of the largest differences between the two reference states.
This difference highlights how the choice of chemical potential can affect the strength of the adsorbate-substrate interaction. 
A plot showing the variation of the adsorption energy with the chemical potential referenced to that of the bulk metal is provided in Supplementary Figure 7 for MoS$_2$, MoSe$_2$, WS$_2$, and WSe$_2$. 
Referencing the chemical potential to the bulk metal results in more positive (i.e., less exothermic) adsorption energies, and reflects a scenario where the metal adatom is part of a bulk reservoir.
To focus on the differences in adsorption energy arising from the interactions between the monolayer and adsorbate, we continue our discussion using the chemical potential of the isolated adatom.

\subsection{Rationalizing trends in adsorption energy}
\label{sec_dband_ads}

To rationalize the observed trends, we employ the \textit{d}-band center model, which combines the effects of electron filling and the formation of bonding and anti-bonding orbitals between the adsorbate and the substrate (also see Section~\ref{sec_dband_method}).

Physically, the strength of interaction between the adsorbate and substrate is related to the filling of the anti-bonding states between the re-normalized valence states of the substrate and adsorbate upon adsorption. 
Because anti-bonding states generally lie energetically above the electronic band edge, the center of the \textit{d}-states is an indicator of the bond strength between adsorbate and substrate.
The center of the \textit{d}-band is often referenced to the Fermi level, which determines the energy level of half occupation and thus delineates the filled and unfilled electronic states.  
Thus, a higher in energy \textit{d}-band center relative to the Fermi level is indicative of a stronger adsorption interaction, as fewer anti-bonding states are occupied.
Hence, the \textit{d}-band model provides a basis for explaining differences and trends in adsorption energetics. 

Traditionally, this model focuses on the position of the \textit{d}-band center of the transition metal \textit{substrate} relative to the Fermi level as the typical adsorbates are \textit{organic chemical intermediates involved in catalytic reactions.}
By contrast, the adsorbates in this study are transition metal adatoms on select substrates (MoS$_2$, MoSe$_2$, WS$_2$, WSe$_2$) that show minimal difference in their structural relaxations (Figure~\ref{figure_lat_param}).
In this study, we apply the \textit{d}-band center model by focusing on the \textit{d}-band centers of the transition metal \textit{adsorbates} instead of the TMD substrates.
This adaptation of the \textit{d}-band center model leverages the underlying principle that the orbital filling and the position of the \textit{d}-band center describe differences in adsorption energies.
This dependence on the electronic structure is illustrated in Equations~\ref{dbc_equation} and \ref{effective_dbc_equation}.
Studies such as Qiao \textit{et al}.~\cite{qiao_vertical-orbital_2021} have also investigated hydrogen adsorption on single-atom-anchored 2D basal planes, finding a correlation between the $d_{z^2}$-band center and adsorption strength. 
These observations collectively suggest that similar principles may govern interactions involving transition metal adatoms on low-dimensional materials.

\begin{figure}[H] 
    \centering
    \includegraphics[width=0.8\textwidth]{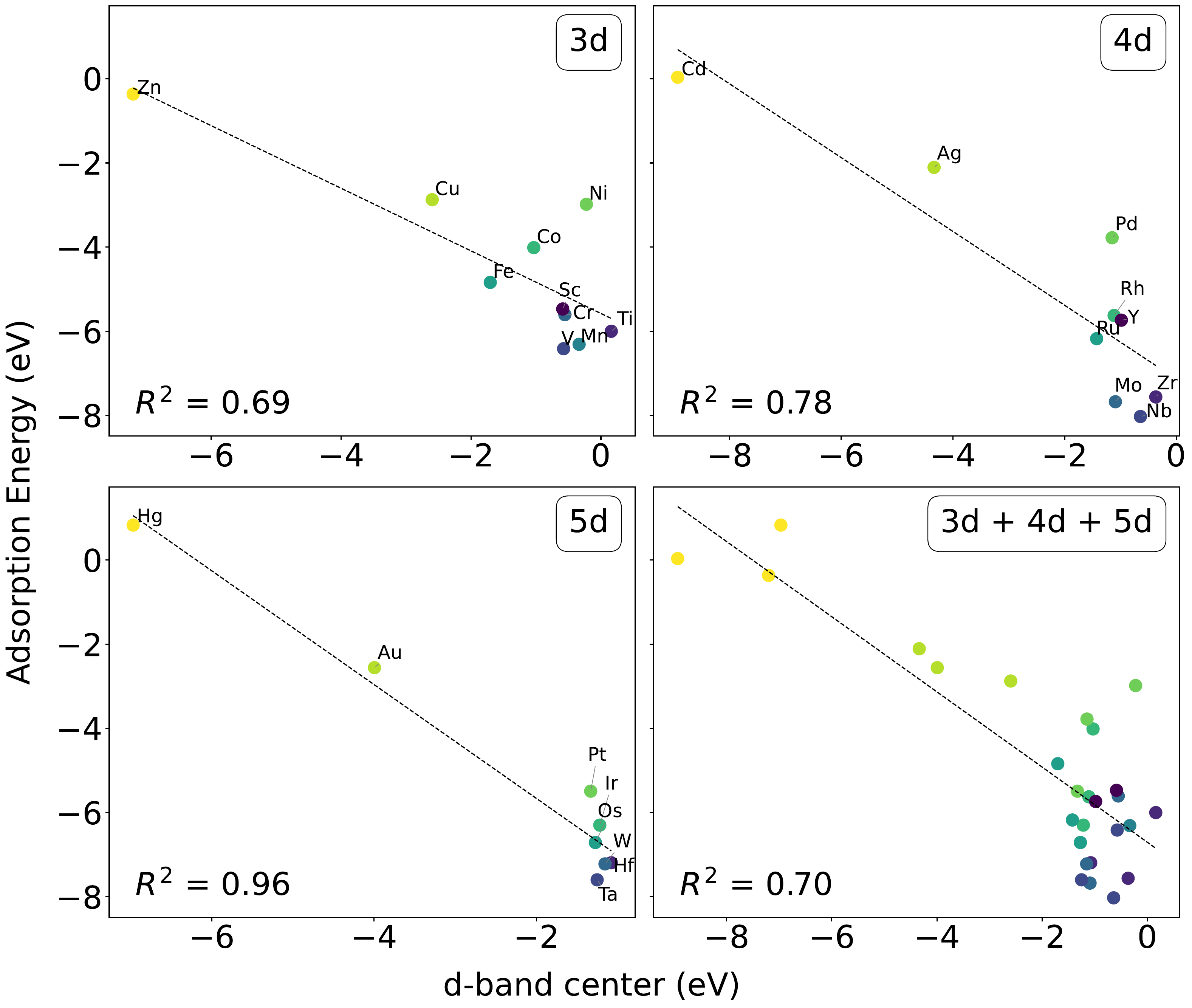}
    \caption{Comparison of the \textit{d}-band centers of the adsorbates, referenced to the Fermi energy, with their respective adsorption energies on MoS$_2$. A \textit{d}-band center of zero indicates alignment with the Fermi energy. The plots are categorized by specific \textit{d}-block elements ((a) 3\textit{d}, (b) 4\textit{d}, (c) 4\textit{d}, (d) all together) and color-coded by periodic groups. Linear fits with accompanying R$^2$-values are provided as a guide the eye.}
    \label{figure_MoS_2_d_band_split}
\end{figure}

As observed from Figure~\ref{figure_MoS_2_d_band_split} for MoS$_2$, the early transition metals, such as Hf, Zr, and Ti, have \textit{d}-band centers shifted closest to the Fermi level when compared to other transition metals. 
A similar observation can be made for MoSe$_2$, WS$_2$, and WSe$_2$ (see Supplementary Figures 8, 9 , and 10). 
According to the \textit{d}-band model, this shift of the \textit{d}-bands closer to the Fermi level correctly anticipates a stronger adsorption energy, as the anti-bonding states would be minimally occupied. 
In contrast, the late transition metals, such as Cd, Hg, Zn, Au, Cu, and Ag, have \textit{d}-band centers shifted farther from the Fermi level, indicating lower energy levels for the anti-bonding orbitals, leading to weaker adsorption energy. 
Consequently, the \textit{d}-band centers for the re-normalized electronic states of the middle transition metals are shifted by an intermediate amount, resulting in intermediate adsorption energies and strengths. 
By comparing the shifts of the \textit{d}-band centers of different adsorbates on the surfaces of TMDs, we can directly correlate the electronic structure of the adsorbates with their adsorption energies. 
This approach provides a clear understanding of the trends observed across different transition metal adsorbates, highlighting the utility of the \textit{d}-band center model in rationalizing trends in adsorption properties in resistive switching contexts~\cite{hammerWhyGoldNoblest1995}.

Interestingly, we observe that by separating the transition metal adsorbates into 3\textit{d}, 4\textit{d}, and 5\textit{d} metals, there is a more robust correlation between the computed adsorption energy and the \textit{d}-band center (Figure~\ref{figure_MoS_2_d_band_split}). 
We hypothesize that this difference in correlation based on period stems from the increasing spatial extents of the 3\textit{d}, 4\textit{d}, and 5\textit{d} orbitals.
An examination of the hybridization characteristics of the adsorbate reveals no discernible trends (see Supplementary Figures 11 and 12).
Deviations from the linear \textit{d}-band model are known, such as when multiple transition metal species are involved as in skin alloys~\cite{xin_communications_2010}. 
The multiplicity in spin state of transition metals has also been shown to break linearity in adsorption energetics~\cite{moltved_dioxygen_2020}, particularly for the spatially-localized, highly-correlated 3\textit{d} metals where higher-accuracy wavefunction-based methods are often required to capture and resolve different spin states~\cite{moltved_dioxygen_2020,ioannidis_towards_2015}.
Supplementary Figures 8, 9, and 10 provide equivalent data for MoSe$_2$, WS$_2$, and WSe$_2$.

\begin{figure}[!ht] 
    \centering
    \includegraphics[width=0.85\textwidth]{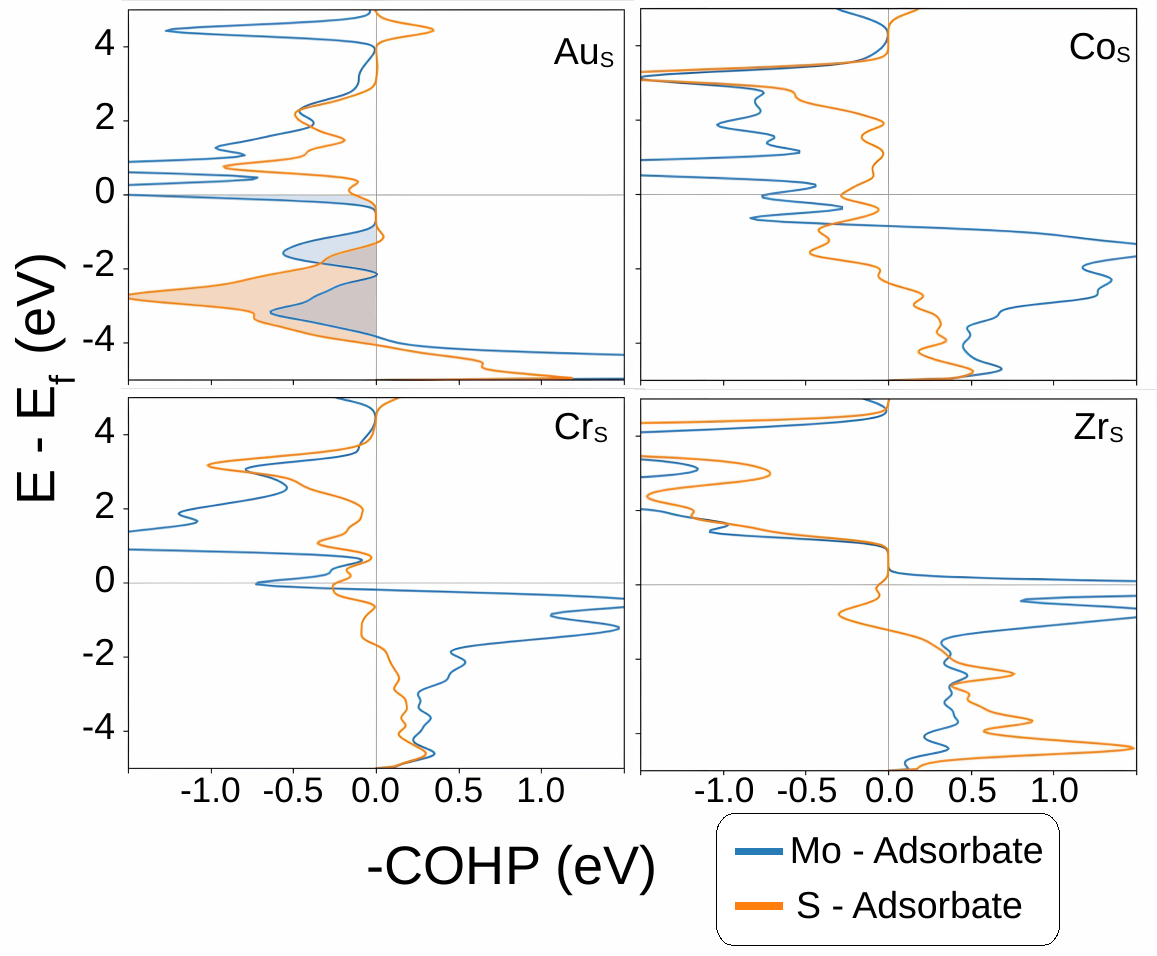}
    \caption{Crystal Orbital Hamilton Population (COHP) plots of representative transition metal adsorbates from groups 11 (Au), 9 (Co), 6 (Cr), 4 (Zr) on MoS$_2$ are presented. The blue plots show the interactions between the transition metal adsorbate and the molybdenum adsorption site in the monolayer; the orange plots show the interactions between the transition metal adsorbate and the surrounding sulfur in the monolayer. Positive COHP values correspond to bonding interactions, whereas negative values indicate anti-bonding interactions. An estimate of the percent anti-bonding character up to the Fermi level is computed for each adsorbate and corresponds to anti-bonding interactions between the adsorbate and the MoS$_2$ atoms. The highlighted region for Au$_S$ is used to illustrate the portion of the -COHP curves used to compute the percent anti-bonding character. A similar method is used for each of the different adsorbates. The areas under the bonding and anti-bonding regions provide insight into the overall strength of the interaction, with higher bonding contributions correlating to stronger adsorption. The gold, cobalt, chromium, and zirconium adsorbates exhibit  74.18\%, 45.87\%, 29.17\%, and 9.62\% anti-bonding character with sulfur (orange), respectively.}
    \label{figure_some_lobster}
\end{figure}

We next turn to crystal orbital Hamilton population (COHP) analysis.
COHP analysis provides insights into the bonding and anti-bonding character in the interactions between adsorbates and substrates, helping to further understand the nature of adsorption. 
By decomposing the electronic density of states (DOS) into its bonding and anti-bonding components as a function of energy, we recover trends in line with the \textit{d}-band center model. 
Notably, our COHP analysis reveals that anti-bonding interactions involving sulfur atoms play a role in the adsorption energy trends.
Figure~\ref{figure_some_lobster} shows examples for how the degree of anti-bonding character is computed based on COHP analysis.
From this analysis, we observe that an increasing contribution of anti-bonding interactions from molybdenum and sulfur leads to a destabilization of adsorption energies. 

\begin{figure}[H] 
    \centering
    \includegraphics[width=0.75\linewidth]{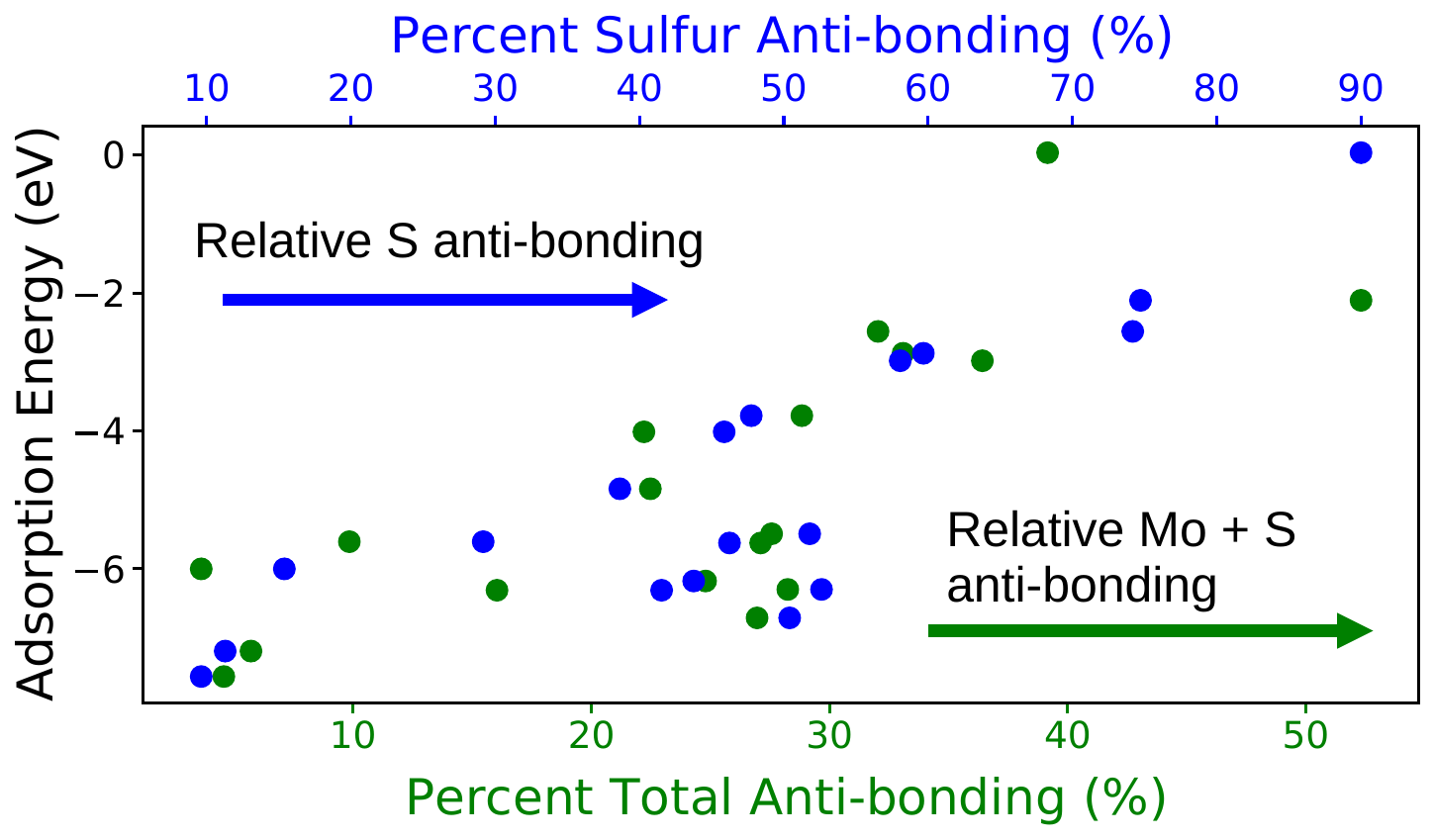}
    \caption{Comparison of COHP results describing the percentage of anti-bonding interactions of the adsorbate and sulfur (ab-ADS-S) and total anti-bonding interactions in MoS$_2$. The anti-bonding interaction percentage between the adsorbate and sulfur is correlated with adsorption energy with an R$^2$ value of 0.70, while the combined anti-bonding character has an R$^2$ value of 0.57.}
    \label{figure_lobster_S}
\end{figure}

We used our COHP analysis to decompose the overall bonding and anti-bonding interactions by species. 
Figure~\ref{figure_lobster_S} shows that percent sulfur anti-bonding character and adsorption energy are correlated ($R^2 \sim$  0.70).
This analysis is in line with the \textit{d}-band model, in which the percent anti-bonding character of states below the Fermi level correlates with inversely with the adsorption strength, and further reveals the species contributions. 
For additional examples, the COHP analysis shows that Hf, Zr, and Ti have the lowest anti-bonding interactions with their nearest neighbor sulfur atoms (with percent anti-bonding contributions of 11.28\%, 9.62\%, and 15.39\%, respectively), corresponding to generally stronger adsorption.
Hg, Cd, and Zn exhibit the greatest anti-bonding interactions with their nearest neighbor sulfur atoms (with percent anti-bonding contributions of 84.12\%, 90.0\%, and 85.21\%, respectively), corresponding to generally weaker adsorption. 

These findings highlight that adsorption strength is governed by multiple factors, including charge redistribution and antibonding interactions. 
A comparison of the correlation between adsorption energy with Bader charge, percent Mo anti-bonding character, and percent S anti-bonding character is provided in Supplementary Figure 13. 
This finding also provides physical insights as to the selection of descriptors for random forest regression of adsorption energies in the work by Frey \textit{et al}. \cite{frey_machine_2020}, where it was not clear why the mean number of \textit{p}-valence electrons was ranked as the highest in importance.  
Here, our analysis reveals that the charge loss or gain from the chalcogen atoms, whose valence electrons are primarily of \textit{p}-character, strongly influences the filling of anti-bonding orbitals, correlating with adsorption strength.  

We also conducted an analysis of the hybridization character of the adsorbates' \textit{d}-orbitals to evaluate whether differences in adsorption energy observed across different periods could be better understood. 
For adsorbates within the same group, despite their general similarity in electron configuration, the hybridization of the 3\textit{d}, 4\textit{d}, and 5\textit{d} orbitals varies with the substrate. 
To quantify the amount of $d_{xy}, d_{xz}, d_{yz}, d_{z^2}$, and $d_{x^2-y^2}$ orbital character, we integrate the projected density of states over a fixed energy range, from -2.0 eV to 0.5 eV relative to the Fermi level, to capture any mid-gap states introduced by the adsorbates; these data may be found in Supplementary Figures 11 and 12. This range was chosen to focus on the energy window most relevant to resistive switching and charge transport, where adsorbate-induced hybridization with defect sites is expected to be strongest. It includes the entire band gap and portions of the conduction and valence bands, ensuring contributions from both filled and empty hybrid states are accounted for. While deeper integration windows would include broader bonding states, we found that the relative orbital contributions remain largely consistent across different windows for most adsorbates.

The variation in adsorption energies across different adsorbates have some correlation with the relative \(d\)-orbital contributions. 
For example, in the case of the group 4 adsorbates (Ti, Zr, Hf) on the MoS$_2$ substrate, it can be observed that Ti exhibits the weakest adsorption energy of -6.13 eV. 
The $d$-orbital contributions of Ti are relatively evenly distributed among $d_{xy}$, $d_{yz}$, $d_{xz}$, and to a lesser extent $d_{z^2}$, with moderate contributions from $d_{xz}$ and $d_{yz}$. 
In contrast, Zr and Hf show a more pronounced contribution from the $d_{yz}$ and $d_{xz}$ orbitals, which may facilitate more effective spatial alignment with the MoS$_2$ surface.
Similarly, in group 6 metals, Mo exhibits the strongest adsorption energy of -7.78 eV on MoS$_2$ and similarly has a larger contribution of the $d_{z^2}$ orbital. 

\begin{figure}[!ht] 
    \centering
    \includegraphics[width=1\linewidth]{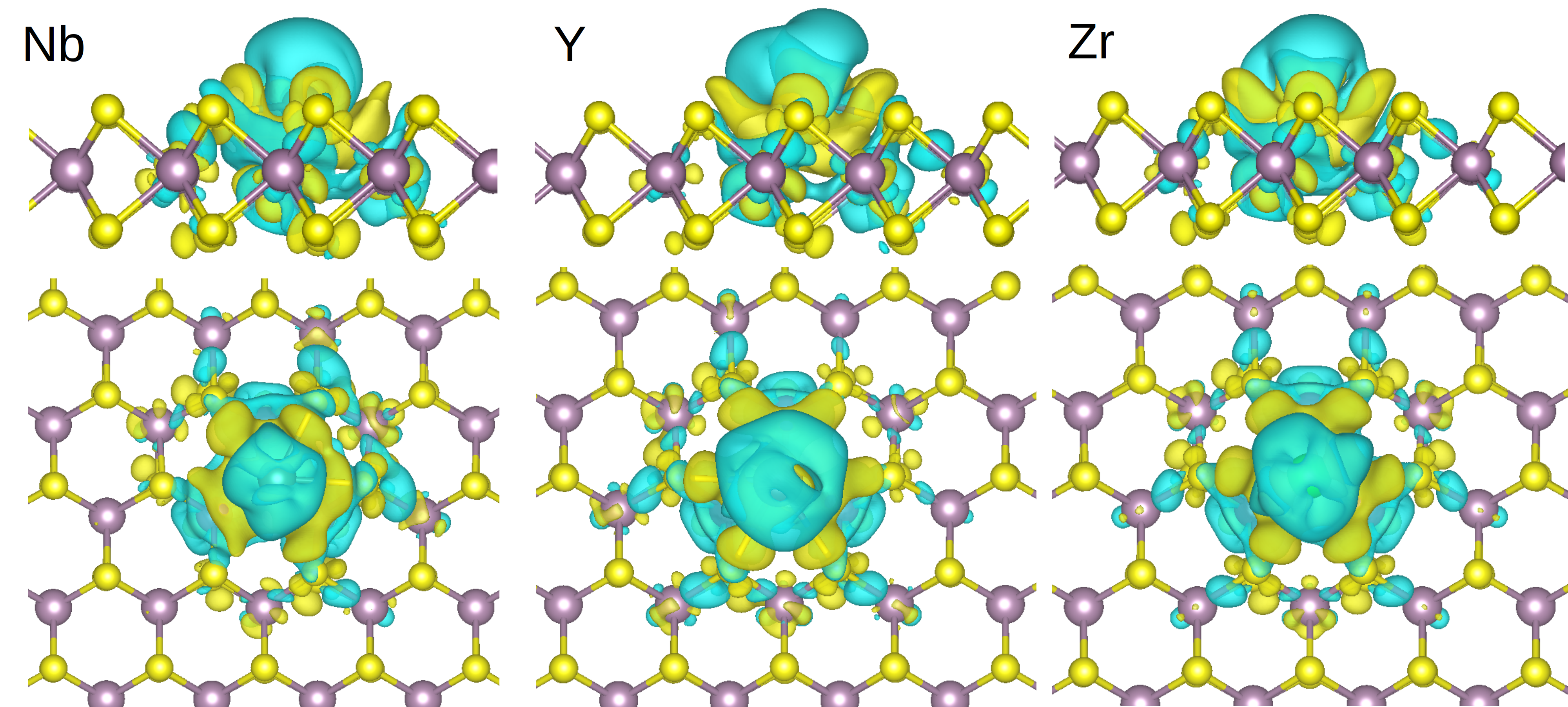}
    \caption{The charge density difference between the adsorbed structure, the v$_S$ MoS$_2$ structure, and the adsorbate plots of Nb, Y, and Zr, viewed along the b-axis and the c-axis. The yellow distribution corresponds to charge accumulation and the blue corresponds to charge depletion. The isosurface value is equivalent to 0.0015 e/\AA$^{3}$}
    \label{CHGDIFF_Nb_Zr_Y}
\end{figure}

Charge density difference (CDD) plots provide further insight into the nature of these interactions. For instance, Nb, Y, and Zr are shown in Figure~\ref{CHGDIFF_Nb_Zr_Y}, which all show a higher degree of $d_{yz}$ and $d_{xz}$ contributions.
The CDD plots reveal regions of charge depletion near the adsorbate and charge accumulation near the substrate. 
This visual evidence supports the argument that these adsorbates have ionic bonding characteristics, where the spatial alignment of the $d_{yz}$ and $d_{xz}$ orbitals of the adsorbate enables greater interaction with the substrate.

Overall, the variation in adsorption energies within the same group can be partially understood to arise from differences in the hybridization and spatial orientation of specific $d$-orbitals with the substrate orbitals. 

\subsection{Prediction of trends in adsorption energies}
\label{sec_SISSO}
To explore the prediction of adsorption energies, we employ the Sure Independence Screening and Sparsifying Operator (SISSO) \cite{ouyang_sisso_2018}, a compressed sensing technique to predict adsorption energies using basic descriptors. 
In this context, dimensionality refers to the number of features or descriptors used to represent the data. 
High-dimensional datasets, where the number of potential descriptors is large, can be challenging because they increase the complexity of the model; there is also a risk of overfitting, in which the model becomes too tailored to the specific dataset and loses its ability to generalize to new data.

SISSO addresses these challenges by first reducing the dimensionality of the feature space discarding weakly correlated features through sure independence screening (SIS), focusing the analysis on the most relevant descriptors. 
This reduction is crucial for managing complexity and ensuring that the model remains both robust and generalizable.
Next, the sparsifying operator (SO) identifies the optimal $n$-dimensional descriptor by selecting a minimal set of features that best predict the adsorption energies, ensuring that the model is both accurate and interpretable. 
Unlike many machine-learning techniques, SISSO can effectively perform non-linear regression on small datasets while preserving interpretability, making it ideal for complex datasets where transparency is essential.
A list of the initial set of descriptors used in the SISSO analysis is provided in the SI.

We leverage a dataset of readily available atomic properties for the substituents in each structure to construct the SISSO descriptor \cite{ouyang_sisso_2018, tranAnisotropic2019}.
In Section~\ref{sec_ads_trends}, we demonstrated the significance of bonding and anti-bonding contributions and adsorbate charge loss or gain to adsorption energies. 
This leads us to consider whether the \textit{d}-band center calculation, which typically requires a detailed density of states (DOS) analysis, could be avoided with proxy descriptors. 
Identifying an alternative method or descriptor that bypasses this step could improve the computational efficiency of adsorption energy predictions.

To assess the performance of our SISSO-based approach, we employed a leave-one-out (LOO) validation strategy. 
In this approach, one TMD substrate was entirely excluded from the training set and used as an independent test case,  repeating the process for (MoS$_2$, MoSe$_2$, WS$_2$, and WSe$_2$. 
The complexity of our SISSO models is governed by user-defined parameters, such as descriptor dimensionality and the allowed complexity in combining features (see Supplementary Section 3).

Figure~\ref{figure_sisso_ads} presents the predicted versus reference DFT adsorption energies for the best-performing 5D SISSO model across different LOO cases. 
For instance, when trained on MoSe$_2$, WS$_2$, and WSe$_2$, the model was tested on MoS$_2$ (Figure~\ref{figure_sisso_ads}b and c). 
Supplementary Table 2 quantifies model performance across all LOO scenarios, summarizing training and test set R$^2$, mean squared error (MSE), and mean absolute error (MAE).
The SISSO model maintains test set R$^2$ values ranging from 0.789 to 0.892 across all LOO cases. 
We note that even 1D SISSO models can achieve R$^2$ = 0.64 (Supplementary Figure 18). 
Physical quantities, including electronegativity ($X$), ionization energy ($IE$), work function ($WF$), atomic radius ($R_c$), and valency, were consistently retained in all models.

SISSO models that include the \textit{d}-band center and structural information (e.g., average distance between adsorbate and neighboring molybdenum atoms) were also evaluated (see Supplementary Figures 21 and 22). 
These models provide further improvements in R$^2$ up to 0.91. 
Increasing the dimensionality beyond 5D resulted in diminishing returns. 
Simpler models that explicitly account for some known limitations of the \textit{d}-band center model also improve the descriptive nature of the \textit{d}-band center model. 
Supplementary Figures 19 and 20 show that including the covalent radius information about the atomic species in each system, structural information in the form of average distance between the adsorbate and substrate, and filling information (i.e., number of valence electrons) improves the correlation and descriptive nature of the \textit{d}-band center model. 
In both 2D and 3D SISSO models, utilizing the \textit{d}-band center, structural information, filling information, and categorical descriptors based on periodic blocks, we achieve R$^2$ values of 0.80 and 0.84 (Supplementary Figures 19 and 20), respectively).
Overall, the agreement between SISSO-predicted and DFT reference values (Figure \ref{figure_sisso_ads}) and the relative consistency in R$^2$, MAE, and MSE across LOO test cases (Supplementary Table 2) suggest that atom-based descriptors can be used to predict trends in adsorption energetics in TMDs.

Compared to other computational studies on high-throughput defect and adsorption energy calculations, our error metrics provide insight into the degree of error introduced when using a minimal set of simple descriptors, such as atomic radii and electronegativities, in contrast to more complex and computationally expensive non-linear regression methods.
For example, Mannodi-Kanakkithodi and colleagues \cite{mannodi-kanakkithodi_universal_2022} reported mean squared errors (MSEs) of 0.94–1.05 eV for defect formation energies up to 11 eV in zinc blende materials using advanced machine learning models like random forests and neural networks.
Similarly, Wan \textit{et al}.~\cite{wanDatadrivenMachineLearning2021} observed MSEs of 0.21–0.45 eV for oxygen vacancy formation energies in metal oxides using various multiple linear and machine-learning regression models.
These studies contain regression models that require large amounts of data and have limited interpretability.
Frey \textit{et al}.~\cite{frey_machine_2020} used ensemble random forest regression to achieve R$^2$= 0.74 and MAE = 0.67 in adsorption energies of metal adatoms in TMD monolayers.
The top two out of ten reported descriptors included the mean number of $p$-valence electrons and the chemical potential.
The chemical potential is already a term in the calculation of the adsorption energy, and reveals little of the adsorbate-absorbent interactions.
Our previous COHP analysis reveals that the interactions of the adsorbate with surrounding chalcogen atoms may partially explain the descriptor of mean $p$-valence electrons.

In comparison, the SISSO approach offers reasonable accuracy for simpler models with smaller datasets. 
Here, we choose descriptors that directly link to basic chemical parameters (e.g., atomic radii, electronegativity, ionization energies).
We show that models containing one descriptor (e.g., $d$-band center) to models based on three to five descriptors can explain a majority of the trends in the adsorption energy (R$^2$ = 0.89 to 0.91).
The simplicity of our SISSO models leads to a tradeoff of slightly higher deviations from reference data, as seen in our adsorption energy calculations (Figure~\ref{figure_sisso_ads}a).
Nevertheless, these models serve as a valuable starting point, balancing computational efficiency with model simplicity and reasonable predictive accuracy, particularly in high-throughput studies where efficiency is critical or in resource-limited computational studies.

Finally, we comment on the anticipated transport characteristics of the adsorbed metal adatoms, a pre-requisite for such point defects to be used in resistive switching contexts.
Adsorption of gold onto sulfur vacancies in MoS$_2$ is known to lead to up to around ten orders of magnitude of change in the normalized conductivity, including the out-of-plane component, depending on the strain present~\cite{anvari_nature_2024,ge_library_2021,hus_observation_2021}.
The change in conductivity is accompanied by a partially-filled defect level with predominantly Au $d$-character that imbues a degree of metallic character to the substrate.
Notably, calculations from Anvari and Wang~\cite{anvari_nature_2024} predicted a ratio of around 10 orders of magnitude difference between the low- and high-resistive states, which compares comparably with the experimentally observed ratio of $6-8$ orders of magnitude difference~\cite{hus_observation_2021}.
Hence, the change in conductivity can, to first order, be explained by the increase in density of states near the Fermi level~\cite{anvari_nature_2024}. 
When inspecting the DOS of the adsorbates studied here, we observe that 76\% of the configurations tested are expected to show enhanced metallic-like behavior with the adatom adsorbed (a list of the elements may be found in Supplementary Table 1).
Further investigation of the transport properties is merited and the topic of future study.

\subsection{Physical model for resistive switching based on point defects}
\label{sec_DDA}

Building on the computed adsorption energies, we establish a connection between these energies and the switching energy in vertically-stacked memristors composed of monolayer TMDs.
Our model of the switching energy is based on the dissociation-diffusion-adsorption (DDA) model~\cite{ge_atomristor_2018,ge_library_2021} (Supplementary Figure 1). 

In the DDA model, an adatom from the electrode dissociates diffuses across the surface before adsorbing to a pre-existing anion vacancy (Supplementary Figure 1).
This process involves three key energy components: dissociation energy, diffusion energy, and adsorption energy.

We observe a correlation (R$^2$ = 0.96) between the calculated adsorption energies of various metal adatoms on monolayer TMDs and the experimentally observed set power in corresponding memristor devices (Supplementary Figure 2). 
Specifically, stronger (more negative) adsorption energies are associated with reduced set power. 
This correlation suggests the adsorption energy plays a role in the switching energetics of the device.
The energy of dissociation is determined by the chemical identity of the electrode material and thus is assumed to be constant for a given adatom species.
An additional analysis comparing calculated adsorption energies with the cohesive (bulk) energies of the corresponding metals is shown in Supplementary Figure 3. 
This comparison reveals a strong negative correlation ($R^2 \approx 0.80$), indicating that metals with higher cohesive energies typically exhibit stronger adsorption energies.
A full study of diffusion is outside the scope of this paper, but is the topic of ongoing study.
Thus, our focus from an energetics point of view is the adsorption energy.

The DDA model is supported by previous measurements of single-defect memristors in MoS$_2$ based on local scanning tunneling microscopy images~\cite{hus_observation_2021}. 
In these studies, gold electrodes were observed to dissociate into adatoms, which then interact with the MoS$_2$ surface, modulating the resistive switching state.
Different electrode materials can lead to distinct switching mechanisms. 
For instance, devices utilizing silver electrodes often exhibit filamentary switching behaviors, where the formation and dissolution of conductive filaments play a crucial role in the resistive switching process~\cite{ling_mos2-based_2023, min_investigation_2021, wang_surface_2019}. 
A more systematic comparison across electrode materials is required, for which this study provides initial materials-based insights into expected trends.

As discussed in Section 1 of the Supplementary Information, a correlation between the measured switching energy~\cite{ge_library_2021} and the computed adsorption energies for the common TMDs suggests that the adsorption energy plays an important role in the energetics of resistive switching based on transition metal dichalcogenides.
Previous studies, such as the one by Frey \textit{et al} \cite{frey_machine_2020}, have used a similar idea of using binding energies and formation energies to identify optimal defects for applications in neuromorphic and quantum information processing. 
Frey \textit{et al}.~\cite{frey_machine_2020} used random forest regression to predict adsorption energies and identify potential descriptors but did not fully delve into the defect chemistry that dictated trends in adsorption energy. 
This study emphasizes understanding the adsorption mechanism between 2D TMDs and various transition metal adsorbates, to further elucidate how the valency, periodicity, and electronic configuration of the metal affect  the adsorption energetics.

\section{Discussion}\label{sec_discussion}
We have shown that the adsorption energy is relatively insensitive to the chemical species of the substrate when considering commonly studied TMDs (MoS$_2$, MoSe$_2$, WS$_2$, WSe$_2$).
Saliently, the trends in adsorption species may be explained using the \textit{d}-band model, a model based on electronic structure descriptors most well-known in its application to catalysts. 
Hence, the switching energy is materials-dependent and that dependence can be rationalized. 
An earlier study on the adsorption of Li in transition metal dichalcogenides for battery applications demonstrated that the work function can serve as a simple descriptor for explaining trends in Li adsorption\cite{douLithiumAdsorption2D2020}.
In this study, we generalize this concept to adsorption energies of metal adatoms in low-dimensional materials.
Thus, insights from electrochemistry and catalysis may be effectively cross-applied in resistive switching contexts, and can be applied more broadly to classes of substrates and adsorbates than previously demonstrated.

We note that these defect calculations assumed an isolated monolayer.
In reality, the interactions between the TMD and the electrode at the interface are known to significantly impact the electronic properties of the TMD. 
For example, in our previous study on MoS$_2$/Au heterostructures~\cite{anvari_nature_2024}, the Au electrode induces a strain on the MoS$_2$ and results in a charge transfer that alters the electronic structure and hence the resulting adsorption energies.
Nevertheless, the \textit{d}-band model can similarly be used to explain changes in adsorption energy due to tensile strain~\cite{anvari_nature_2024}.

Additionally, in idealized models, the energetics of adsorption and desorption are often considered symmetric, meaning the energy gained via adsorption is equivalent to the energy required for desorption, and vice versa. However, in real systems where external forces like applied voltage, such as those involving resistive switching in 2D materials, this symmetry often breaks down due to surface defects, strain, and energy barriers. Thus, adsorption energies alone cannot explain the asymmetry observed in the SET/RESET I-V characteristics of measured samples~\cite{ge_library_2021}. We posit that this asymmetry likely requires, at minimum, the combined effects of explicitly including the presence of the electrode and an external potential to capture the full mechanism.
Furthermore, while this study covers the energetics of defect formation, it does not account for the kinetics or dynamics by which defect formation occurs (e.g., diffusion processes). Each of these aspects is anticipated to play a role in the overall resistive switching characteristics of the device, thus meriting further study. While such effects are neglected here, our study provides an important baseline for understanding the intrinsic defect chemistry of the switching layer material.

Overall, we have shown with the simplifications of the dissociation-diffusion-adsorption (DDA) model\cite{ge_library_2021} that performance metrics of the resistive switching, such as the switching energy and the adsorption energy, are materials dependent and can be rationalized based on materials-based descriptors of the electronic structure. 
For structures of 2D materials that differ from the 2H phase considered here, we anticipate incorporating structure-sensitive descriptors will further improve the model~\cite{andersenScalingRelationsDescription2019}. 
A study on the generalizability of models for adsorption energy beyond transition metal dichalcogenides or to multiple adsorption events is merited and a topic of further study.

Within and beyond resistive switching, defects in low-dimensional materials play a prominent role in next generation computing devices. 
Indeed, two-dimensional (2D) materials are ideal platforms for hosting quantum defects~\cite{linDefectEngineeringTwodimensional2016,mcdonnellDefectsTransitionMetal2022}. 
An important implication of our study in rationalizing adsorption energies of point defect complexes in TMDs is the identification of electronic structure and atomic descriptors for both the substrate and electrode, which extends our ability to engineering designer defects for targeted applications.
For practical device applications, we expect that the adsorption energy of transition metal adatoms plays a role in the switching energy (see also Supplementary Figure 2). 
That is, a large (negative) adsorption energy corresponds to a greater stability of the adsorbate, which may be important for long-term operation. 
On the other hand, a low (near zero) adsorption energy corresponds to a low switching energy, which may contribute to more energy-efficient resistive switching.

As a practical recommendation, gold is typically chosen for its inertness to the chemical environment~\cite{hammerWhyGoldNoblest1995}. 
Our study demonstrates that this inertness extends to adsorbed adatoms, as gold exhibits a lower adsorption energy, making it favorable for resistive switching applications where energy-efficient and reversible switching is desired. 
Based on our calculations, we find that copper (Cu) and nickel (Ni), which are CMOS-compatible metals, have relatively low adsorption energies (-2.96 eV and -4.01 eV, respectively), which may make them suitable for switching applications with lower energy consumption.
In contrast, metals like tungsten (W) and molybdenum (Mo) exhibit large adsorption energies (-7.78 eV), which may make them suitable for applications where retention of the adsorbate is needed. 
These findings are particularly relevant for CMOS integration, as the ability to tune adsorption energies through material selection can provide control over both the stability and switching energy of the device.
Finally, we anticipate these models and descriptors to be useful in screening novel materials defects in databases~\cite{haastrupComputational2DMaterials2018,bertoldoQuantumPointDefects2022} and high-throughput calculations towards expanding our materials repertoire of low-dimensional materials for the next generation of electronic devices.

Ultimately, we investigated the adsorption energies of various transition metal adsorbates on different monolayer TMD substrates, utilizing both computational methods and theoretical models to understand the underlying trends. 
We evaluated the adsorption energy of a metal adatom on a pre-existing anion vacancy for MoS$_2$, MoSe$_2$, WS$_2$, and WSe$_2$ and suggest these adsorption energies as proxy for investigating resistive switching energy. 
The adsorption energies displayed significant variation depending on the periodic group of the transition metal, with early transition metals (Hf, Zr, Ti) exhibiting the strongest adsorption energies, and late transition metals (Cu, Au, Ag, Zn, Cd, Hg) showing the weakest, while the middle transition metals exhibited moderate adsorption energies in between the early and late transition metals. 
We demonstrated that these trends in the adsorption energy could be largely captured by trends in the \textit{d}-band center with respect to the Fermi level, and as a descriptor can capture most of the trend in adsorption energies.
COHP analysis provided further insights into the nature of bonding and anti-bonding interactions between the adsorbates and their nearest neighbor sulfur atoms in the TMDs, emphasizing the importance of the local environment surrounding the adsorption site in determining the adsorption energy. 
We used this understanding to perform compressed sensing on atom-based descriptors, which does not require a DFT calculation. 
This SISSO analysis demonstrated promise in predicting absolute adsorption energies, using compound descriptors based off of the ionization energy, electronegativity, and atomic radii of the adsorbate and adsorption site.

Our findings highlight the importance of considering both the electronic structure and the local bonding environment in understanding materials-based trends in the adsorption behavior on TMD substrates to a greater degree than previously shown for transition metal adsorbate species. 
While the prediction of absolute adsorption defect energetics is on-going, we emphasize the understanding trends already provides materials design criteria for low-dimensional materials in energy-efficient resistive switching applications.
Notably, this study provides a simplified materials-driven framework for modeling and understanding adsorption energy and resistive switching, achieved through cross-applying principles of the \textit{d}-band model from electrocatalysis on adsorption mechanisms and energetics.
By integrating multiple theoretical frameworks, including the \textit{d}-band center model, bond analysis using COHP, and SISSO, we have presented a valuable tool for materials-driven design for future memristive devices.

\section{Methods}
\label{sec_methods}

\subsection{Computational Parameters}
\label{sec_comp_param}

The calculations in this work were performed using the Vienna \textit{Ab initio} Simulation Package (VASP 6.3.0) \cite{kresse_ab_1993, kresse_efficiency_1996, kresse_efficient_1996} with the projector augmented-wave (PAW) method and the Perdew-Burke-Ernzerhof (PBE) exchange-correlation functional \cite{kresse_ultrasoft_1999, perdew_generalized_1996}. 
A kinetic energy cut-off of 400 eV was used for the plane-wave basis set expansion. 
The Brillouin zone was sampled using a 2 $\times$ 2 mesh, and spin polarization was taken into account, as it is expected to be significant when dealing with transition metal adsorbates \cite{bhattacharjee_improved_2016}.
Convergence criteria were set to an energy difference of $10^{-5}$ eV for electronic steps and force difference of 10 meV/\r{A}.
For density of states calculations, a $k$-point grid of 4 $\times$ 4 (equivalent to a resolution of 0.07 \r{A}$^{-1}$) was used, resulting in a difference less than 0.01 eV in Fermi energies).

To investigate the overall relationship between the structure and composition of a TMD-adatom complex, we use the adsorption energy defined by Eq.~\ref{adsorptionE_equation}. 
Here, $E_{ads}$ is the adsorption energy, which represents the energy required for the adsorption of a single adatom onto the substrate. 
The terms $E^{tot}_{ads/sub}$, $E^{tot}_{sub}$, and $E^{tot}_{ads}$ denote the total energies of the structure with the adsorbate, the substrate without the adsorbate, and the chemical potential of the adatom, respectively. 

\begin{equation} \label{adsorptionE_equation}
    E_{ads}=E^{tot}_{ads/sub}-E^{tot}_{sub}-E^{tot}_{ads}
\end{equation}

\noindent In general, the adsorption energy depends on the chemical potential of the adatom, and hence depends on the bath from which the adatom originates, which in turn is determined by experimental conditions.
We explore the effect of the chemical potential, ranging from the isolated adatom to that referenced to the bulk metal, on the adsorption energy in Section~\ref{sec_ads_trends}. 
In Eq.~\ref{adsorptionE_equation}, we assume the chemical potential to be effectively that of an isolated adatom in order to focus on the adsorption energetics due to chemical and electronic differences in the adsorbate or monolayer.
In general, other factors such as diffusion barriers are expected to influence memristive switching. 
Indeed, prior experimental and theoretical studies, including recent work from Anvari and Wang~\cite{anvari_nature_2024}, have demonstrated that interfacial effects such as strain, interlayer spacing, and associated electronic modifications strongly impact the switching characteristics.  
Nevertheless, adsorption energies reflect energetically favorable initial and final configurations that are consistent with experimentally observed switching states~\cite{hus_observation_2021}.
Thus, our present analysis of adsorption energetics provides a foundation for ongoing and future studies exploring these dynamic processes in greater detail.

To minimize spurious defect-defect interactions, a supercell size of 8 × 8 (192 atoms) was used with a vacuum size of 20~\r{A}. 
The relaxed structures shown in Figure~\ref{figure_lat_param} illustrate the distance between adjacent molybdenum atoms near the defect site. 
We focus on the adsorption of a metal adatom on a pre-existing chalcogen vacancy as the representative point defect complex, as our prior work has shown this configuration to be more energetically favorable than the adsorption of the adatom on a pristine monolayer~\cite{anvari_nature_2024}.
For example, adsorption energies of Ag and Au on pristine MoS$_2$ were calculated to be -0.59 eV and -0.64 eV, respectively, at the site directly above the molybdenum, with similar energies within 0.2 eV at other sites (-0.74 eV above the chalcogen site, -0.73 at the hollow site). 
In contrast, the adsorption energy at the chalcogen vacancy site was calculated to be -2.56 eV for gold. 
This is consistent with the dissociation-diffusion-adsorption (DDA) model~\cite{ge_library_2021} presented in Section~\ref{sec_DDA}, where the mechanism involves the dissociation of the adatom from the electrode, diffusion across the TMD, and eventual adsorption at an anion vacancy site.

\subsection{The \textit{d}-band model}
\label{sec_dband_method}
The \textit{d}-band center model is an important theoretical framework used to understand the interaction between adsorbates and transition metal surfaces in catalysis applications \cite{hammer_electronic_1995,norskov_density_2011,norskovElectronicFactorsCatalysis1991,norskovUniversalityHeterogeneousCatalysis2002}, including those involving transition metal dichalcogenides (TMDs) \cite{tsaiTheoreticalInsightsHydrogen2015,tsaiUnderstandingReactivityLayered2014,hongHowDopedMoS2016}. 
Originally, this model was used to study the catalytic activity of bulk transition metal surfaces with adsorbates such as hydrogen \cite{hammer_electronic_1995}, nitrogen \cite{skulason_theoretical_2011}, and carbon dioxide \cite{peterson_activity_2012}.

Here, we apply the \textit{d}-band model to rationalize trends in adsorption energy as they relate to resistive switching applications. 
The \textit{d}-band model is based on the Newns-Anderson model \cite{newns_self-consistent_1969,anderson_localized_1961}, describes the interaction between an adsorbate's energy levels and the electronic structure of the substrate. This model provides insight into how the position and width of the substrate's \textit{d}-band influence adsorption strength and chemical reactivity.
Subsequent refinements to this model have enhanced its accuracy and applicability, such as the inclusion of spin~\cite{bhattacharjee_improved_2016} and the inclusion of structural information via compound descriptors~\cite{andersenScalingRelationsDescription2019}, and we utilize aspects of these refinements in our study.
This approach utilizes the energy level of the \textit{d}-band center and the shift induced by the adsorbate as key descriptors for adsorption energy. 

The \textit{d}-band center, $\varepsilon_d$, is defined as the first moment of the density of states (DOS) of the $d$-electrons and can be calculated as:

\begin{equation} \varepsilon_d = \frac{\int_{-\infty}^{\infty} \varepsilon \, n(\varepsilon) \, d\varepsilon}{\int_{-\infty}^{\infty}n(\varepsilon) \, d\varepsilon} \label{dbc_equation}\end{equation}

\noindent where $\varepsilon$ represents the energy eigenvalues, and $n(\varepsilon)$ is the density of states of the bands with projected \textit{d}-character.
In our calculations, we define the \textit{d}-band center over a fixed energy range that includes both regions below and above the Fermi level between the different adsorbates.
This approach allows for a direct comparison of \textit{d}-band centers across different adsorbates.

In this study, we select an energy range of \(-10\, \mathrm{eV}\) to \(1.5\, \mathrm{eV}\), referencing \(E_F = 0\, \mathrm{eV}\), to ensure that the \textit{d}-band contributions both below and above the Fermi level are adequately captured. 
This range is chosen to ensure consistent comparison across different substrate and adsorbate configurations. 
To verify our choice of energy range, we computed the d‑band center using slightly extended and reduced windows (e.g., -11 eV to 1.5 eV and -10 eV to 2 eV) and observed that the calculated \textit{d}‑band center values remained stable, confirming that our selected range reliably captures the relevant \textit{d}‑state contributions.

In our spin-polarized calculations, the \textit{d}-band centers for the spin channels are combined into a single descriptor using the method developed by Bhattacharjee \textit{et al}., which employs a two-centered \textit{d}-band model \cite{bhattacharjee_improved_2016, wang_vaspkit_2021}. 
In this model, a new descriptor called the \textit{effective \textit{d}-band center} is introduced, as shown in Equation \ref{effective_dbc_equation}:

\begin{equation} \label{effective_dbc_equation} \varepsilon^{\text{eff}} = \frac{\sum_\sigma f_\sigma \varepsilon_{d\sigma}}{\sum_\sigma f_\sigma} - (\varepsilon_{d\downarrow} - \varepsilon_{d\uparrow})\mu \end{equation}

\noindent Here, $f_\sigma$ represents the fractional occupancy of spin state $\sigma$, and $\varepsilon_{d\sigma}$ is the conventional Hammer-Nørskov \textit{d}-band center for spin state $\sigma$. 
The final term, $\mu$, is the reduced fractional moment, defined as $\mu = \frac{f_\uparrow - f_\downarrow}{f_\uparrow + f_\downarrow}$. 
The first term in the equation gives the spin-averaged \textit{d}-band center, where the spin-up and spin-down channels are averaged; the second term introduces a shift in the \textit{d}-band center that depends on the spin polarization and the individual spin-resolved \textit{d}-band centers.
The orbital occupation and shift of the \textit{d}-band center will be the two main parameters of focus in this study.
Other parameters in the \textit{d}-band model include the orbital occupation, the coupling between the electronic states of the adsorbate and substrate, and the energy associated with the orthogonalization of the states between the adsorbate and substrate. 
These descriptors are largely based on the electronic structure and thus suggest a generality of the \textit{d}-band model beyond catalysis, i.e., for transition metal adsorbates, extending to applications such as resistive switching based on the formation of point defect complexes. 

As a complement to the \textit{d}-band center model, we also utilize crystal orbital Hamilton population analysis (COHP) to further study the bonding nature of the adsorbate on the various TMDs~\cite{dronskowski_crystal_1993}. These calculations are performed using the Local-Orbital Basis Suite Towards Electronic-Structure Reconstruction (LOBSTER) package~\cite{deringer_crystal_2011, maintz_analytic_2013, maintz_lobster_2016, maintz_efficient_2016, nelson_span_2020}. Crystal Orbital Hamilton Population (COHP) analysis distinguishes between bonding, non-bonding, and antibonding interactions between adjacent atoms by analyzing the projected density of states (DOS) weighted by Hamiltonian matrix elements. Specifically, the COHP between atomic orbitals $\phi_{\mu}$ and $\phi_{\nu}$ at energy $E$ is calculated as follows:

\[
\mathrm{COHP}_{\mu\nu}(E) = H_{\mu\nu}\sum_{j} c_{\mu j} c_{\nu j}\,\delta(E - \epsilon_j)
\]

Here, $H_{\mu\nu}$ is the Hamiltonian matrix element between orbitals $\mu$ and $\nu$, $c_{\mu j}$ and $c_{\nu j}$ are the expansion coefficients of atomic orbitals in eigenstate $j$, and $\epsilon_j$ is the eigenvalue corresponding to eigenstate $j$. The sign of the Hamiltonian-weighted projected DOS (positive or negative) indicates the nature of the interactions: negative COHP values represent bonding (energy-lowering, stabilizing) interactions, whereas positive COHP values represent antibonding (energy-raising, destabilizing) interactions.

To quantify the bonding and antibonding interactions, we integrate the COHP over the energy range from $-\infty$ to the Fermi level ($E_F$). The \textit{antibonding} contribution corresponds to the integral over the positive part of COHP ($\mathrm{COHP}^{+}$):

\[
\text{Antibonding contribution} = \int_{-\infty}^{E_F} \mathrm{COHP}^{+}(E)\,dE
\]

Similarly, the \textit{bonding} contribution corresponds to the integral over the negative part of COHP ($\mathrm{COHP}^{-}$):

\[
\text{Bonding contribution} = \int_{-\infty}^{E_F} \mathrm{COHP}^{-}(E)\,dE
\]

The percent antibonding character is then calculated as:

\[
\text{Percent antibonding (\%)} = \frac{\int_{-\infty}^{E_F} \mathrm{COHP}^{+}(E)\,dE}{\int_{-\infty}^{E_F} \left|\mathrm{COHP}(E)\right|\,dE} \times 100\%
\]

This method provides a detailed understanding of chemical bonding by explicitly linking the DOS with the underlying electronic interactions dictated by the Hamiltonian, enabling a quantitative distinction between bonding and antibonding states across the relevant electronic structure of the system.

Finally, we use the Sure Independence Screening and Sparsifying Operator (SISSO) \cite{ouyang_sisso_2018} to develop interpretable descriptors that leverage chemical intuition to predict adsorption energy (see Supplementary Table 3 in the SI for the complete list of descriptors). 
SISSO is a symbolic regression method that systematically combines chemically relevant primary features—such as ionization energy, electronegativity, and atomic radii—using mathematical operations to generate a vast space of potential descriptors. 
By creating a large pool of candidate features and applying a sparsifying operator, SISSO identifies a minimal set of the most predictive and interpretable descriptors, effectively capturing nonlinear relationships between input features and adsorption energy. 
This approach enables models that are physically meaningful and avoids overfitting by selecting descriptors that generalize well across different datasets. 
The strength of SISSO lies in its ability to reduce the problem's dimensionality, distilling complex material behaviors into a few key descriptors. 
This enhances predictive accuracy and provides significant insights into the determinants of adsorption energy.
Further details are provided in Supplementary Information (SI).

\backmatter

\section*{Data Availability}

The main data supporting the findings of this study are provided within the paper and Supplementary material. Additional data is openly available on Zenodo at 
\hyperlink{https://doi.org/10.5281/zenodo.13892108.}{DOI: 10.5281/zenodo.13892108}, Ref \cite{lee_understanding_2024}

\section*{Acknowledgments}
This research was primarily supported by the National Science Foundation through the Center for Dynamics and Control of Materials: an NSF MRSEC under Cooperative Agreement No. DMR-2308817. 
The authors acknowledge the Texas Advanced Computing Center (TACC) at The University of Texas at Austin for providing computational resources that have contributed to the research results reported within this paper. 
D.A acknowledges the support of the National Science Foundation (NSF) award \#2422934, and the Office of Naval Research (ONR) grant N00014-24-1-2080.
\section*{Author Contributions}
\noindent B.H.L:  Investigation; Data collection; Formal analysis; Visualization; Writing - original draft; Writing - review \& editing. 
J.F.: Data collection; Writing - review \& editing. 
D.A.: Conceptualization; Writing - review \& editing (supporting). 
W.W.: Conceptualization; Resources; Formal analysis; Investigation; Methodology; Writing - original draft; Writing - review \& editing; Funding acquisition (lead); Project Administration; Supervision. 
\section*{Competing Interests}
Authors B.H.L, J.F, W.W declare no financial or non-financial competing interests. Author D.A serves as an Editorial Board Member of this journal and had no role in the peer-review or decision to publish this manuscript. Author D.A declares no financial competing interests.

\bibliography{bibliography}

\section*{Figures}\label{sec_figures}

\begin{figure}[H] 
   \centering 
   \includegraphics[width=1\textwidth]{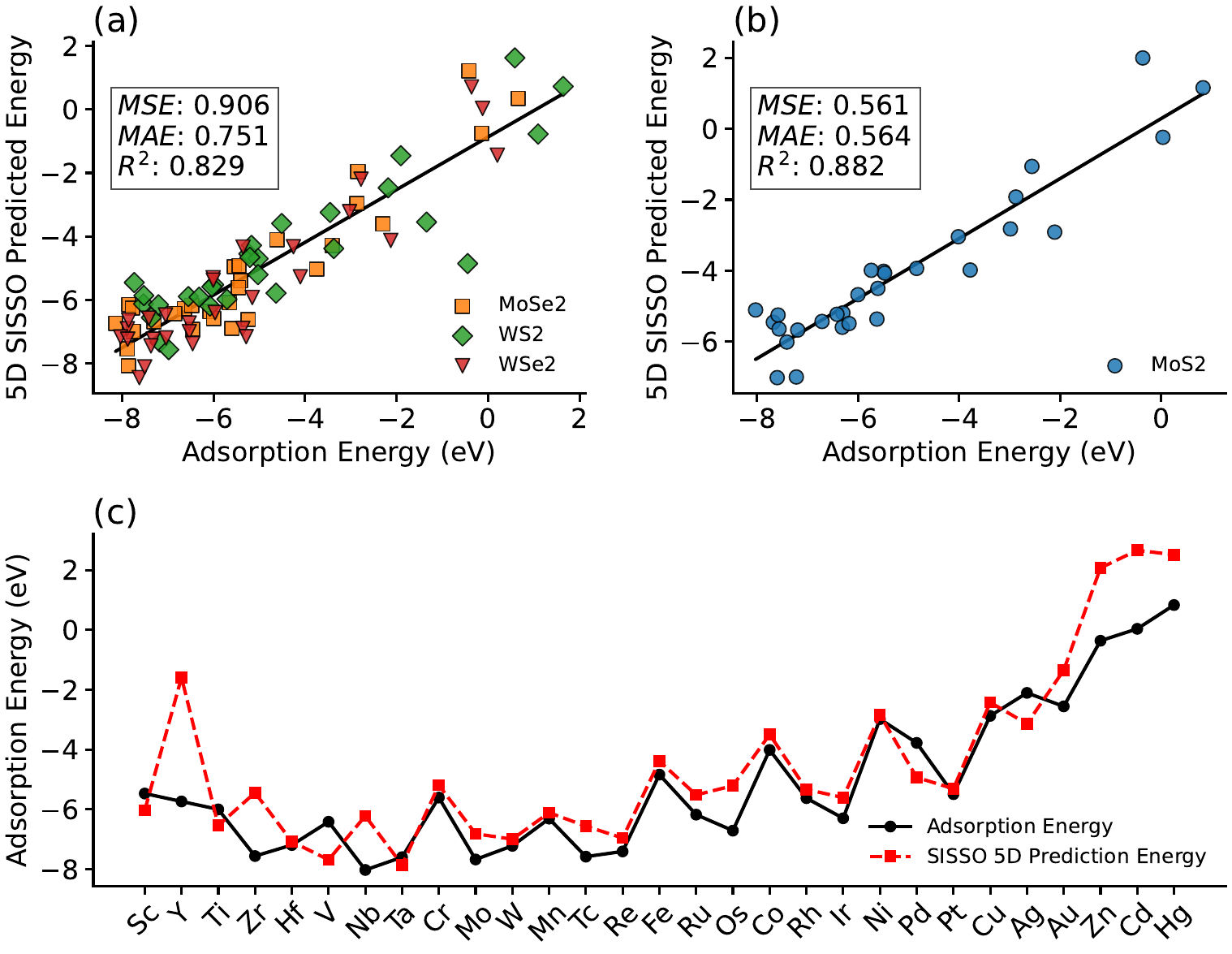} 
   \caption{Prediction of adsorption energies using the best-performing 5D SISSO descriptor. (a) The model was trained on MoSe$_2$, WS$_2$, and WSe$_2$ data. (b) The trained model was tested on MoS$_2$. (c) Comparison of reference DFT values (black) and SISSO-predicted values (colored) for MoS$_2$. } \label{figure_sisso_ads} 
\end{figure}

\begin{figure}[H] 
    \centering
    \includegraphics[width=0.75\textwidth]{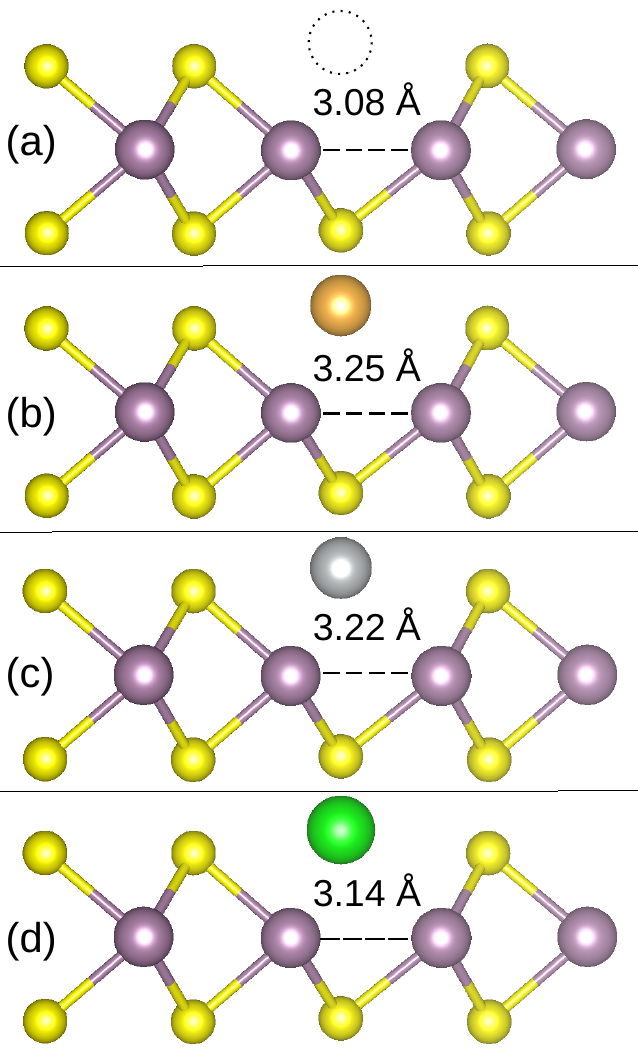}
    \caption{Relaxed atomic structures of MoS$_2$ with (a) a single sulfur vacancy, (b) gold adsorbate, (c) platinum adsorbate, (d) zirconium adsorbate. The distances between the nearest neighbor molybdenum atoms in each structure are indicated.}
    \label{figure_lat_param}
\end{figure}

\end{document}